\documentstyle[aps,preprint,epsf,psfig]{revtex}

\begin{document}

\draft
\title{Solving the Crisis in Big-Bang Nucleosynthesis
by the Radiative Decay of an Exotic Particle}
\author
{Erich~Holtmann$^{a,b}$, Masahiro~Kawasaki$^c$ 
and Takeo~Moroi$^b$}
\address{
$^a$Department of Physics, University of California, 
Berkeley, CA 94720\\
$^b$Theoretical Physics Group, Lawrence Berkeley Laboratory, 
University of California, Berkeley, CA 94720\\
$^c$Institute for Cosmic Ray Research, University of Tokyo, Tokyo 
188, JAPAN}

\date{\today}

\maketitle

\begin{abstract}

We discuss a new mechanism which can solve the crisis in standard
big-bang nucleosynthesis. A long-lived particle $X$ ($10^4{\rm
sec}\lesssim\tau_X\lesssim 10^6{\rm sec}$) which decays into photon(s)
will induce cascade photons, and destroy significant amounts of D and
$^3$He without destroying $^4$He or too much $^7$Li. We numerically
investigate this process and derive a constraint on the properties of
$X$ such that the theoretical values of the primordial light-element
abundances agree with observation. We also present some candidates for
the unstable particle $X$.

\end{abstract}

\newpage

\renewcommand{\thepage}{\arabic{page}}
\setcounter{page}{1}

\renewcommand{\thefootnote}{\arabic{footnote}}
\setcounter{footnote}{0}

The abundances of the light elements are a crucial test of
the big-bang nucleosynthesis (BBN) scenario, since BBN
precisely predicts the primordial ({\it i.e.}, before stellar processing)
abundances of D, $^3$He, $^4$He, and
$^7$Li~\cite{BBN-review}. In the past,
improvements in the theoretical and
observational estimates of the light elements seemed to give us a
better confirmation of the standard BBN scenario.

Recently, however, observations seem to conflict with the theoretical
predictions of standard BBN: if we determine the baryon-to-photon
ratio $\eta$ by using the observed $^4$He abundance, then the
primordial abundances of D and $^3$He predicted from the standard BBN
calculation are too large.  In particular, based on a statistical
argument, Hata {\it et al.} claimed that the standard BBN scenario is
inconsistent with observations at a very high confidence
level~\cite{Ohio-Penn}. Furthermore, Copi {\it et al.} pointed out
that the systematic error in the $^4$He observation has likely been
underestimated if the {\it standard} BBN scenario is correct. These
arguments suggest the need for some modification of the theoretical or
observational estimations of the primordial abundances.

At present, there are several attempts to solve the discrepancy,
though none of them has been confirmed. One way is to adopt some
modification of the observed abundance. In particular, if we assume an
extra systematic error in the estimation of the $^4$He abundance or if
we trust the higher measurements of D in the primordial ${\rm H_I}$ cloud,
then we will have better agreement between theory and
observation~\cite{PRL75-3981}. Alternatively, the models of chemical
and stellar evolution of the light elements may need some change. The
other way to solve the problem is to modify the standard scenario of
BBN. For example, one can reduce the $^4$He abundance by allowing
degenerate electron neutrinos~\cite{NPB372-494}, or by assuming a
massive unstable neutrino with mass about 1~MeV and lifetime about
1~sec~\cite{NPB419-105}. Since BBN, together with the Hubble expansion
of the universe and the cosmic microwave background radiation (CMBR),
is one of the most essential predictions of the big-bang cosmology, it
is important to clarify the origin of the discrepancy.

In this letter, we propose another new mechanism to solve the crisis
by reducing the predicted D and $^3$He abundance. Essentially, if some
exotic particle $X$ decays into photons well after BBN, then the
predictions of standard BBN can be modified. Conventionally, such
radiatively decaying particles have been regarded as sources of
cosmological difficulties~\cite{Lindley,gravitino,KM1,KM2}. However,
we can actually have a {\it better} scenario of BBN if the
cascade photons can destroy significant amounts of D and $^3$He
without affecting $^4$He and $^7$Li. Therefore, in this letter, we derive a
constraint on the properties of $X$ which makes the BBN scenario
viable. We also suggest several candidates for the unstable particle
$X$ based on realistic particle physics models.

First of all, we briefly review the effects of the
radiative decay of $X$. Once a high energy photon is emitted into the
thermal bath, cascade processes are induced.
The high energy photon, electron, and positron spectra are formed
accordingly. Cascade photons then induce photodissociation of
the light elements.
Thus, if the radiative decay of $X$ induces many
cascade photons at a time $t$ after BBN starts ($t\gtrsim
1{\rm sec}$), then the predictions of standard BBN can be modified.
In the photon cascade process, double photon pair
creation ($\gamma +\gamma_{\rm BG}\rightarrow e^{+}+e^{-}$,
with $\gamma_{\rm BG}$ being the photon in thermal background) plays
an important role: almost all the photons with energy larger than
$\epsilon_{\rm thr}\sim m_e^2/22T$ scatter off photons in the
thermal background and become $e^{+}e^{-}$ pairs, since the typical
event rate for this process is much larger than that of other
processes. On the other hand, for photons with energy smaller than the
threshold energy $\epsilon_{\rm thr}\sim m_e^2/22T$, this process
is extremely suppressed by kinematics~\cite{E_thr,KM1,KM2}.
As a result,
radiative decay of $X$ may cause photodissociation of light elements
only if the decay of $X$ occurs at $T\lesssim m_e^2/22Q$, with $Q$
being the threshold energy of photodissociation.  For
example, photodissociation of $^4$He, whose binding energy is about
20MeV, can be effective only for $T\lesssim 500{\rm eV}$, {\it i.e.}
for $t\gtrsim O(10^{6-7}{\rm sec})$.

The effects of the radiative decay of $X$ essentially depend on the
following parameters: the lifetime $\tau_X$ of $X$, the number density
of $X$, and the energy $\epsilon_{\gamma 0}$ of the photons emitted in
$X$ decay.  In this letter, for simplicity, we assume that
$X\rightarrow\gamma +\gamma$ with a $100\%$ branching ratio, so
$\epsilon_{\gamma 0}=m_X/2$. (For the effect of the hadronic decay of
$X$, see Ref.~\cite{hadron}.)  To parameterize the abundance of $X$,
we use the yield variable $Y_X$, which is the ratio of the number
density of $X$ to that of photons: $Y_X=n_X/n_\gamma$. The yield
variable is essentially the number of $X$ in a comoving volume, and it
evolves with time as $Y_X=Y_{X0}e^{-t/\tau_X}$.  Therefore, once
$\tau_X$, $Y_{X0}$, $\epsilon_{\gamma 0}$, and $\eta$ are fixed, we
can calculate the primordial abundances of the light elements.

The procedure used in this letter is as follows. We first solve the
Boltzmann equations for the distribution functions of the high energy
photons and electrons in order to determine the abundance of cascade
photons.  Then, combining the derived cascade spectrum with the
modified Kawano code~\cite{kawano} in which effects of the
photodissociation are taken into account, we calculate the primordial
abundances of the light elements, $y_{\rm 2p}$, $y_{\rm
3p}$, and $Y_{\rm p}$, where $y_{\rm 2p}$ and $y_{\rm 3p}$ are the
primordial number fraction of D and $^3$He relative to hydrogen H,
respectively, and $Y_{\rm p}$ is the primordial mass fraction of
$^4$He. (For details, see Refs.~\cite{KM1,KM2}.)

These theoretical predictions should be
compared with the constraints obtained from observations. First, let
us discuss the constraints on D and $^3$He. Since D is only destroyed
after BBN, the mass fraction of D decreases with time,
and hence~\cite{D-He3}
\begin{eqnarray}
y_{\rm 2,ism}   &\leq& y_{\rm 2p} R_X,
\label{y2ism} \\
y_{2\odot}  &\leq& y_{\rm 2p} R_X,
\label{y2sun}
\end{eqnarray}
where $R_X=X_{\rm H,p}/X_{\rm H,now}$,
$y_2$ is the number fraction of D
relative to H, $X_{\rm H}$ is the mass fraction of H, and the 
subscripts ``ism'', ``$\odot$'', and ``now'' denote
the abundances in interstellar matter, in
the solar system, and in the present universe, respectively.
Furthermore, taking into account the chemical evolution of D
and $^3$He, we obtain a third constraint~\cite{D-He3}:
\begin{eqnarray}
\left\{ -y_{\rm 2p} 
+ \left(\frac{1}{g_3}-1\right)y_{\rm 3p} \right\}y_{2\odot}
- \frac{1}{g_3} y_{\rm 2p} y_{3\odot} 
\nonumber \\
+ \left(y_{\rm 2p}^2 + y_{\rm 2p}y_{\rm 3p}\right) 
R_X
\leq 0,
\label{y2&y3}
\end{eqnarray}
where $y_3$ is the number fraction of $^3$He
relative to H, and $g_3$ is the survival
fraction of $^3$He in stellar processes.
Usually, $g_3$ is estimated to be 0.25 -- 0.5~\cite{g3}.

From observation, the present abundances of the light
elements (with 1-$\sigma$ error) are known to be~\cite{y23_obs}
\begin{eqnarray}
y_{\rm 2,ism} &=& (1.6\pm 0.2)\times 10^{-5},
\label{const_y2ism} \\
y_{2\odot} &=& (2.57\pm 0.92)\times 10^{-5},
\label{const_y2sun} \\
y_{3\odot} &=& (1.52\pm 0.34)\times 10^{-5},
\label{const_y3sun} \\
R_X &=& 1.1 \pm 0.04.
\label{const_RX}
\end{eqnarray}
Here, we briefly comment on the detection of D in the primordial ${\rm
H_I}$ cloud. The absorbed line observed in Ref.~\cite{nature368-599}
suggests $y_{\rm 2}\simeq 2.5\times 10^{-4}$, which is much larger
than the value given in (\ref{const_y2ism}) and (\ref{const_y2sun}).
However, it is claimed that the observed absorption line may be due
to a Doppler-shifted hydrogen line~\cite{9405038}. Therefore, in this paper,
we do not adopt the D abundance obtained in Ref.~\cite{nature368-599}.

For fixed values of $y_{\rm 2p}$, $y_{\rm 3p}$, and $g_3$, by regarding
$y_{\rm 2,ism}$, $y_{2\odot}$, $y_{3\odot}$, and $R_X$
as statistical variables which obey Gaussian
distributions, we can calculate the probability $P(y_{\rm 2p},y_{\rm
3p};g_3)$ that the conditions (\ref{y2ism}) -- (\ref{y2&y3}) are
satisfied simultaneously:
\begin{eqnarray}
P(y_{\rm 2p},y_{\rm 3p}; g_3) =
\int_{\bf V} dy_{\rm 2,ism} dy_{2\odot} dy_{3\odot} dR_X
\nonumber \\
\times f(y_{\rm 2,ism};\bar{y}_{\rm 2,ism},\sigma^2_{y_{\rm 2,ism}})
f(y_{2\odot};\bar{y}_{2\odot},\sigma^2_{y_{2\odot}})
\nonumber \\
\times f(y_{3\odot};\bar{y}_{3\odot},\sigma^2_{y_{3\odot}})
f(R_X; \bar{R}_X, \sigma^2_{R_X}).
\label{prob}
\end{eqnarray}
In Eq.(\ref{prob}), the integration is performed in the volume {\bf
V} in which all the constraints (\ref{y2ism}) -- (\ref{y2&y3}) are
satisfied, and $f(x;\bar{x},\sigma^2)$ is the Gaussian distribution
function for $x$ with mean $\bar{x}$ and variance $\sigma^2$.

By using the probability $P(y_{\rm 2p},y_{\rm 3p};g_3)$, we can obtain
a constraint on the primordial abundance of D and $^3$He.
For a fixed value of $g_3$, we exclude the parameter set
$(y_{\rm 2p},y_{\rm 3p})$ if $P(y_{\rm 2p},y_{\rm 3p};g_3)<0.05$.
Here, we note that our $P(y_{\rm 2p},y_{\rm 3p};g_3)=0.05$
contour with $g_3=0.25$ is almost the same as the 95\%~C.L. constraint
of Hata {\it et al.}
in Ref.~\cite{D-He3}. That is, comparing the constraints on
$y_{\rm 3p}$ for a fixed value of $y_{\rm 2p}$, we checked that the
discrepancy between the
two approaches is at most $(10-20)\%$; hence, we
conclude that they are consistent.

Next, we quote the constraint on the primordial
abundance of $^4$He, as estimated from the low-metallicity ${\rm
H_{II}}$ regions~\cite{He4}:
\begin{eqnarray}
Y_{\rm p} = 0.232 \pm 0.003 ({\rm stat}) \pm 0.005 ({\rm syst}).
\end{eqnarray}

\begin{figure}[p]
\epsfxsize=13cm
\centerline{\epsfbox{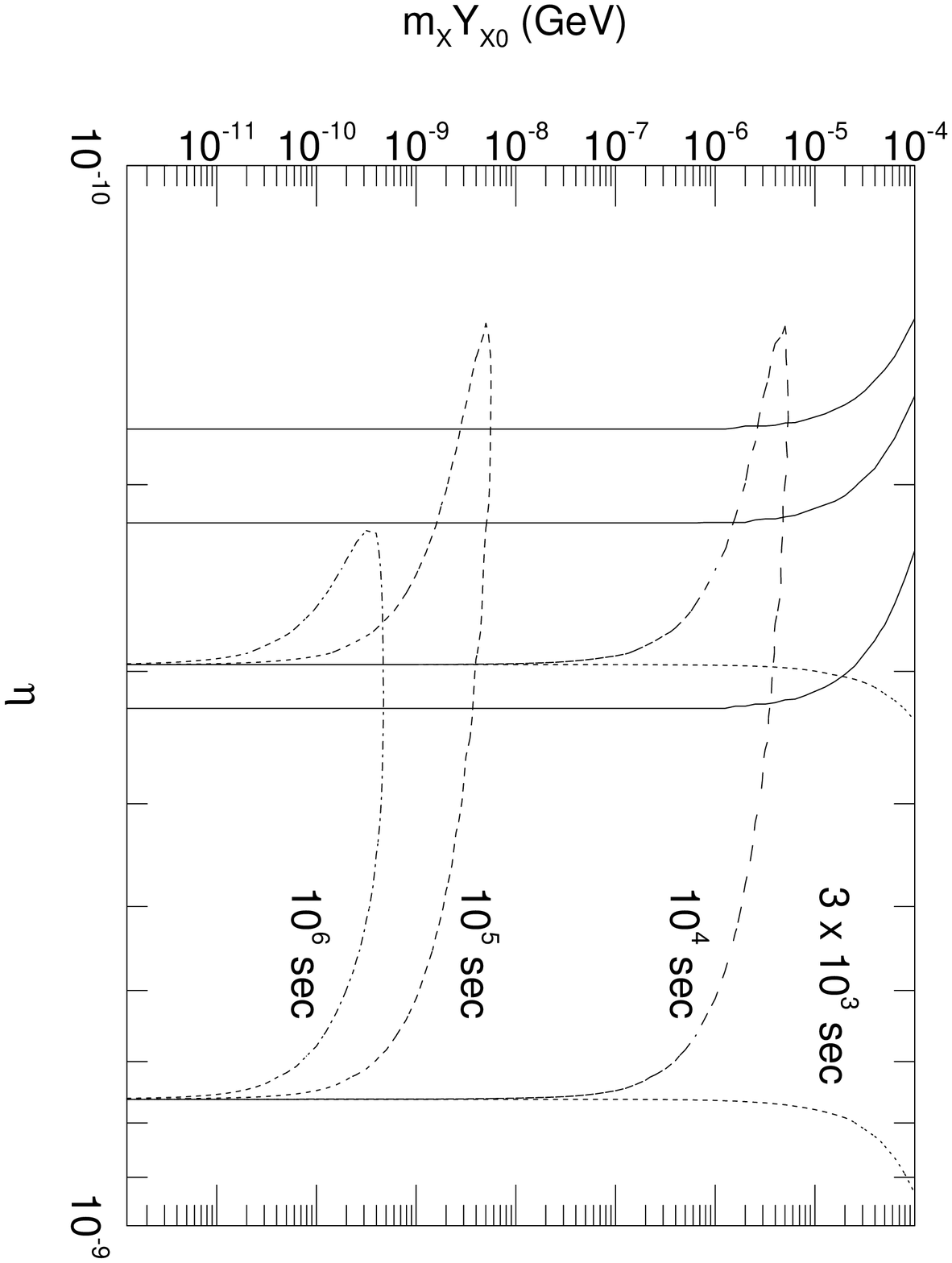}}
\caption{Constraints from $^3$He and D, $P(y_{\rm 2p},y_{\rm 3p};g_3)
\geq 0.05$, shown in the $\eta$ vs. $m_XY_{X0}$ plane for $m_X=1{\rm
TeV}$, and $g_3=0.25$.  The lifetime of $X$ is take to be $3\times
10^3$sec (dotted line), $10^4$sec (long-dashed line), $10^5$sec
(short-dashed line), and $10^6$sec (dotted-dashed line).  The allowed
regions are the interiors of the curves.  The solid lines are the
contours which yield $Y_{\rm p}=0.232, 0.235,$ and $0.240$ (from left
to right) for $\tau_X =3\times 10^3{\rm sec}$.}
\label{fig:1tev}
\end{figure}

We are now ready to show our numerical results. In our analysis, we
compare the constraint on the baryon-to-photon ratio $\eta$ derived
from (\ref{y2ism}) -- (\ref{y2&y3}), with the contours which yield the
correct primordial abundance of $^4$He.  In Fig.~\ref{fig:1tev}, we
show the contour for the 5\% probability constraints in the
$m_XY_{X0}$ vs. $\eta$ plane for several values of $\tau_X$. Here, we
take the mass of $X$ to be $1{\rm TeV}$, a neutron lifetime of
887.0sec~\cite{PDG}, and $g_3=0.25$. Also, we show the contours which
will yield specified abundances of $^4$He.

The typical behavior of the constraint can be understood as follows.
If $\epsilon_{\gamma 0}Y_X$ is too small, then the abundance of
cascade photons is so suppressed that the standard BBN scenario is
almost unchanged.  In this case, constraints from D and $^3$He prefer
$\eta$ to be $(3-7)\times 10^{-10}$, which is too large to be
consistent with the $^4$He constraint.  However, if $\epsilon_{\gamma
0}Y_X$ has the correct value, then sufficient amounts of D and $^3$He
are destroyed by cascade photons, and a smaller value of $\eta$ is
allowed. (Remember that, in the standard BBN scenario, smaller values
of $\eta$ give larger values of $y_{\rm 2p}$ and $y_{\rm 3p}$.)  In
particular, as can be seen in
Fig.~\ref{fig:1tev}, $Y_{\rm p}=0.232$ may be consistent with the
constraints from D and $^3$He for $g_3=0.25$.  Furthermore, we also
checked that, even for $g_3=0.5$, $Y_{\rm p}=0.235$ is acceptable if
$\epsilon_{\gamma 0}\sim 10{\rm MeV}$. Thus, if there is an exotic
particle, we can have a better agreement between observation and the
theoretical prediction. If $\epsilon_{\gamma 0}Y_{X0}$ is too large,
then too much D and $^3$He are destroyed; hence such a parameter
region is excluded. Notice that the distribution function of cascade
photons depends primarily on the total amount of injected
energy~\cite{KM2} if the mass of $X$ is much larger than
$\epsilon_{\rm thr}$. In this case, our results are almost independent
of $m_X$ for fixed $m_XY_{X0}$.

One may worry that too much $^7$Li may be photodissociated in this
scenario, since the threshold for the photodissociation of $^7$Li is
very low ($\sim 2.5{\rm MeV}$). However, the cross section for the
photodissociation of $^7$Li is about four times smaller than that of D
dissociation, so the abundance of $^7$Li is not reduced much compared
with the D abundance. Numerically, we checked that (30--40)\% of
$^7$Li is photodissociated in our scenario. For example, for $\eta
=2\times 10^{-10}$, we get $n_{\rm ^7Li}/n_{\rm H}=(0.7-1.0)\times
10^{-10}$, and a larger abundance of $^7$Li is obtained if we use a
larger or smaller value of $\eta$.  (Notice that we prefer
$\eta\lesssim 3\times 10^{-10}$. For this range, $^7$Li production
though $^7$Be is not important. For $\eta\gtrsim 3\times 10^{-10}$, we
cannot predict $^7$Li abundance since the cross section for $^7$Be
photofission is not available.)  Taking into account the uncertainty
in the observed abundance of $^7$Li, {\it viz.} $n_{\rm ^7Li}/n_{\rm
H}=(1.4^{+2.1}_{-0.7})\times 10^{-10}$~\cite{CopiSci}, the $^7$Li
abundance in our scenario is consistent with observation.
Furthermore, in our scenario $^6$Li can be produced via ${\rm
^7Li(\gamma, {\it n})^6Li}$. This reaction may produce much more
$^6$Li than standard BBN does. Numerically, we expect a few \% of
$^7$Li can be converted into $^6$Li for $\tau_X\lesssim 10^{6}$sec.
This is consistent with present observation~\cite{6Li} ($n_{\rm
^6Li}/n_{\rm ^7Li}\lesssim 0.1$), and may be checked in future
observations.

\begin{figure}[p]
\epsfxsize=13cm
\centerline{\epsfbox{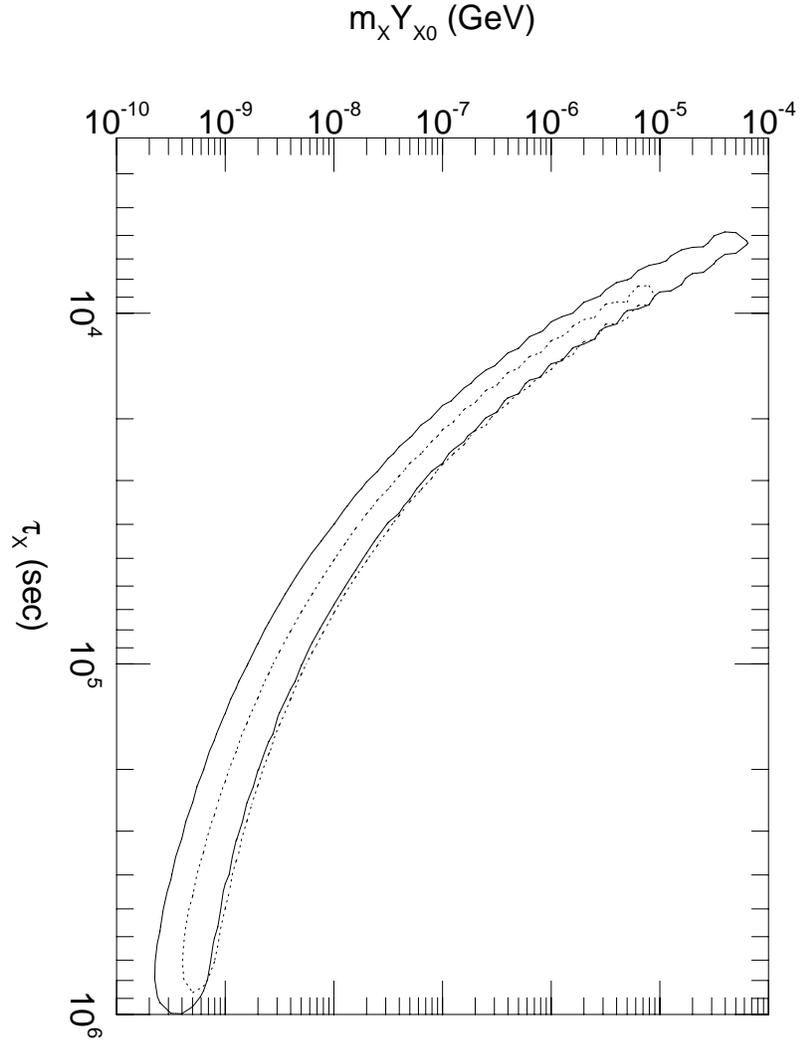}}
\caption {The contour for $P(y_{\rm 2p},y_{\rm 3p};g_3)=0.05$ with $m_X=1{\rm
TeV}$ and $g_3=0.25$. $\eta$ is determined so that $Y_{\rm p}=0.232$
(dotted) and 0.235 (solid).  Notice that the interiors of the contours are
favored.}
\label{fig:t-my}
\end{figure}

Let us now discuss the constraints on $\tau_X$. First of all, the
blackbody spectrum of CMBR observed by COBE~\cite{APJ420-439} gives us
a severe constraint on particles with lifetime longer than $\sim
10^6{\rm sec}$~\cite{PRL70-2661}, which is when the double Compton
process ($\gamma +e^-\rightarrow\gamma +\gamma +e^-$) freezes out.  In
particular, if $\epsilon_{\gamma 0}Y_{X0}\sim 10^{-9}{\rm GeV}$, then
$\tau_X\gtrsim 10^6{\rm sec}$ is forbidden~\cite{PRL70-2661}.
Furthermore, if $\tau_X\gtrsim 10^6{\rm sec}$ and $\epsilon_{\gamma
0}\gtrsim 20{\rm GeV}$, then cascade photons destroy $^4$He
effectively, which may result in the overproduction of D and $^3$He.
Thus, we have an upper bound on $\tau_X$ of $\sim 10^6{\rm sec}$. On
the other hand, as $\tau_X$ decreases, the threshold for double photon
pair creation $\epsilon_{\rm thr}$ becomes low, and the number of
photons contributing to D and $^3$He photodissociation decreases, for
a fixed value of $\epsilon_{\gamma 0}Y_{X0}$.  In this case, a larger
value of $m_XY_{X0}$ is required in order to destroy sufficient
amounts of D and $^3$He, as can be seen in
Fig.~\ref{fig:1tev}.  However, a larger initial mass density of $X$
would speed up the expansion of the universe, so $^4$He may be
overproduced.  These arguments exclude $\tau_X$ shorter than $\sim
10^4{\rm sec}$. As a result, to make our scenario viable, we must
adopt $10^4{\rm sec}\lesssim\tau_X\lesssim 10^6{\rm sec}$.  In
Fig.~\ref{fig:t-my}, we show the allowed region ({\it i.e.}, the
region where $P(y_{\rm 2p},y_{\rm 3p};g_3)\geq 0.05$) on the $\tau_X$
vs. $m_XY_{X0}$ plane.  Here, we use $g_3=0.25$, and we vary $\eta$ so
that $Y_{\rm p}=0.232$ (dotted contour) or 0.235 (solid contour).

Finally, we suggest several candidates for the unstable
particle $X$, specifically within
supergravity~\cite{NPB212-413} models. In such models, there may be several
particles which have a long lifetime ($10^4{\rm
sec}\lesssim\tau_X\lesssim 10^6{\rm sec}$). Probably the most famous
such particle is the gravitino, whose cosmological implications have
been well-investigated~\cite{gravitino,KM1}.  As we will see below,
the gravitino may have the properties required to solve the crisis in
BBN, although in the past it has been regarded as a source of
cosmological difficulties (the so-called gravitino
problem). We can quantitatively check whether the gravitino can solve
the difficulty in BBN, since its interactions are almost unambiguously
determined~\cite{NPB212-413}.  If the gravitino decays only into a
photon and photino pair, then its lifetime is estimated as
\begin{eqnarray} 
\tau_X \simeq 4\times 10^5~{\rm sec} \times (1{\rm TeV}/m_{3/2})^3,
\end{eqnarray}
where $m_{3/2}$ is the gravitino mass. Thus, if the gravitino mass is
about 1TeV, its lifetime is about $10^{5-6}{\rm sec}$, which is
appropriate for our purpose.
Notice that the photino produced by the
gravitino decay interacts very weakly with
the thermal background.
Thus, our results can be applied to the gravitino if
we rescale the $m_XY_{X0}$-axis by a factor of $\sim 0.5$.
Furthermore, assuming an inflationary
universe, the gravitino abundance
is determined just after the reheating period, and 
depends only on the reheat
temperature $T_R$~\cite{KM1}:
\begin{eqnarray}
Y_{X0} \simeq 2\times 10^{-11} (T_R/10^{10}{\rm GeV}).
\end{eqnarray}
For $m_{3/2}\sim 1{\rm TeV}$, $T_R\sim 10^{8-9}{\rm GeV}$ gives
$\epsilon_{\gamma 0}Y_{X0}\sim 10^{-(9-10)}{\rm GeV}$. We note here
that $T_R\sim 10^9{\rm GeV}$ may be realized in the chaotic inflation
model~\cite{PLB129-177} if the inflaton field decays though gravitational
interactions.

If the gravitino is the lightest superparticle, then we can
construct another scenario. In this case, the next-to-the lightest
superparticle (NLSP), which we assume to be the photino, decays into a
gravitino and a photon. By a loose tuning of the parameters, the
NLSP can be identified with an unstable particle $X$ which solves
the difficulty in BBN. Thus, in supergravity models, we
have several candidates for the unstable particle $X$ which can make
the BBN scenario viable.

In summary, an exotic particle with lifetime $\tau_X\sim 10^{4-6}{\rm
sec}$ can
solve the crisis in BBN if $m_X Y_{X0}$
is tuned within a factor of 2--3.
Candidates for $X$ include the
gravitino and the photino, both of
which naturally appear in supergravity models.
Of course, even in other types of models, one may be
able to find a candidate for the radiatively decaying particle $X$.

The authors thank N.~Hata and D.~Thomas for useful discussions and
comments. This work was supported in part by DOE under Contract
DE-AC03-76SF00098 and in part by NSF under grant PHY-90-21139.


%

%
%

\end{document}